\documentstyle[titlepage,12pt]{article}

\def\be{\begin{equation}}
\def\bea{\begin{eqnarray}}
\def\grad{\nabla}
\def\ee{\end{equation}}
\def\eea{\end{eqnarray}}

\def\tmu{TH\epsilon\mu}

\newcount\sectionnumber
\def\sect
{\global\equationnumber=0
\global\advance\sectionnumber by 1
\the\sectionnumber . }
\newcount\equationnumber
\def   \num
{\eqno{\global\advance\equationnumber by 1
\left(\the\sectionnumber .\the\equationnumber \right)}}

\begin{document}
\title{Local Position Invariance and Vacuum Energy Shifts}
\author{C. Alvarez and R.B. Mann \\
Department of Physics \\
University of Waterloo \\
Waterloo, Ontario \\
N2L 3G1}

\date{October 5, 1993\\
WATPHYS-TH93/04}

\maketitle

\begin{abstract}
We discuss tests of the Einstein Equivalence Principle due to energies
which are purely quantum mechanical in origin. In particular, we consider
using Lamb Shift energies to test for possible quantum violations of Local
Position Invariance.
\bigskip

\noindent
({\it to appear in the Proceedings of the 5th Canadian Conference on General
Relativity and Relativistic Astrophysics, University of Waterloo, May 13--
15, 1993})

\end{abstract}

The Einstein Equivalence Principle (EEP) is foundational to our
understanding of gravity. It states that (i) all test bodies fall with the
same acceleration regardless of their composition (the weak equivalence
principle, or WEP) and (ii) all nongravitational laws of physics take on
their special relativistic form. Theories which obey the EEP, such as
general relativity and Brans-Dicke Theory are called metric
theories because they  endow spacetime with a metric $g_{\mu\nu}$ that
couples universally to all non-gravitational fields.
Non-metric theories do not have this feature:  they break universality  by
coupling auxiliary gravitational fields directly to matter. In this
context a violation of the EEP means the breakdown of either
Local  Position Invariance (LPI) or Local Lorentz Invariance (LLI),
so that observers performing local
experiments could detect effects due to their position (if LPI is violated)
or their velocity (if LLI is violated) in an external
gravitational environment by using clocks and rods of differing
composition. Limits on LPI and LLI are set by gravitational red-shift  and
atomic physics experiments respectively \cite{redshift,PLC}, each of which
compares relative frequencies of transitions between particular energy
levels that are sensitive to any potential LPI/LLI-violating effects. The
dominant form of energy governing the transitions in these experiments is
nuclear electrostatic energy, although violations of WEP/EEP due to other
forms of energy have also been estimated \cite{Haug}.

Potential violations of the EEP due to effects which are purely  quantum-
mechanical in origin are not as well understood. We report here on the
results of a calculation which investigates the effects Lamb-shift energies
would have on violations of LPI in the context of a wide class of non-
metric theories of gravity as described by the $\tmu$ formalism \cite{tmu}.
The Lamb shift is the energy shift between the $2s_{1/2}$ and $2p_{1/2}$
states in a Hydrogen-like atom, and is entirely quantum-mechanical in
origin, arising due to interactions of the electron with the quantum
fluctuations of the electromagnetic field \cite{Lamb}.  By considering the
difference between inertial and passive gravitational masses $m_P = m_I +
\Sigma_A \eta_A E_A$ where $\eta_A$ parametrizes the contribution due to
the $A$-th kind of energy, we obtain a crude estimate of the magnitude
of such potential EEP-violating effects to be
\be
\eta \equiv \frac{a_1-a_2}{\frac{1}{2}(a_1+a_2)} \approx
10^{-15}\eta_{Lamb} \label{1}
\ee
where $\eta$ is
the E\"otv\"os ratio \cite{WillBanf} and
$E_A \sim
\frac{4\alpha}{3\pi}\frac{(Z\alpha)^4}{n^3}m_e\ln(\frac{1}{Z\alpha})$
for the Lamb shift. As a comparison, the expected final precision in
the difference between gravitational and inertial mass to be achieved
in the STEP experiment is one part in $10^{17}$ \cite{GR13}.

The $\tmu$ formalism was constructed to study electromagnetically
interacting particles in an external, static, spherically symmetric (SSS)
gravitational field, whose metric tensor is given by $g_{00} = -T(U)$,
$g_{ij} = H(U)\delta_{ij}$, (where $T$ and $H$ are arbitrary functions of
the Newtonian gravitational potential $U = GM/r$) encompassing a wide class
of non-metric (and all metric) gravitational theories. For
the action  of quantum electrodynamics (QED)
with a spinor field $\Psi$ of rest mass $m$ and charge $e$ we have
\begin{eqnarray}\label{2}
I_{QED} &=& \int d^4x det(e)\left[\frac{i}{2}
\left(\overline\Psi \gamma^a e_a^\mu\grad_\mu \Psi -
e_a^\mu(\grad_\mu\overline\Psi)\gamma^a\Psi \right) -m\overline\Psi \Psi
\right] \nonumber\\
& &- \frac{1}{4}\int d^4x \sqrt{-g}g^{\mu\nu}g^{\alpha\beta}F_{\mu\alpha}
F_{\nu\beta}
\end{eqnarray}
where $\grad_\mu = \partial_\mu + \Gamma_\mu - ie A_\mu$ is the covariant
derivative, $A_\mu$ the electromagnetic vector potential with field
strength $F_{\mu\nu} \equiv \partial_\mu A_\nu - \partial_\nu A_\mu$, and
$e_a^\mu$ is the tetrad associated with the metric $g^{\mu\nu} =
\eta^{ab}e_a^\mu e_b^\nu$; $\Gamma_\mu$ is the spin connection. The
$\gamma^a$ are the Dirac matrices, with $\not\!B \equiv \gamma^a e_a^\mu
B_\mu$.

Since we are interested in hydrogen-like atoms,
we can ignore the spatial variation of $T$, $H$,
$\epsilon$, and $\mu$ across the atom and evaluate each of them
at the center of
mass position ${\bf X}=0$. In this context the $\tmu$ formalism modifies
the action (\ref{2}) to $I^\prime_{QED}=I_D+I_{EM}$ where
\be
I_D = \int d^4x'\overline\Psi(i\not\!\partial'+e\not\!\!A'-m')\Psi
\label{3a}
\ee
with $x_\mu '=(c_0x_0,{\bf x})$, $A_\mu '=(A_0/c_0,{\bf A})$, and
$m' = m_eH_0^{1/2}$ and where
\be
I_{EM} = \frac{1}{2}\int(\epsilon_0{\bf E}^2-\frac{{\bf B}^2}{\mu_0})d^4x
+I_{GF}
= -\int d^4\tilde x [\frac{1}{4}\tilde F_{\mu\nu}\tilde F^{\mu\nu}
+\frac{1}{2}(\tilde\partial\cdot\tilde A)^2].
\label{3b}
\ee
with $\tilde F_{\mu\nu}\equiv \tilde\partial_\mu\tilde A_\nu
-\tilde\partial_\nu\tilde A_\mu$, $\tilde x_\mu=(c_*x_0,{\bf x})$,
$\tilde A_{\mu}=\sqrt{{\epsilon}_0c_*}(A_0/c_*,{\bf A})$. Here $I_{GF}$
is a gauge fixing term necessary for quantization and
\be
c_0\equiv {(T_0/H_0)}^{1/2}\qquad c_*\equiv {(\mu_0\epsilon_0)}^{-1/2}
\label{4}
\ee
respectively representing the local limiting speed of massive test bodies and
the local
speed of photons.
In metric theories (which satisfy the EEP)  $\epsilon=\mu={(H/T)}^{1/2}$
\cite{tmu}, implying $c_0=c_*$.

The energy shift of a
state $|n>$ can be expressed as
\begin{equation}\label{6}
\Delta E_n=<n|\Sigma+\Pi-\delta m|n>
\end{equation}
where $\Pi$ and $\Sigma$ are the vacuum polarization and
self-energy contributions, and $\delta m$ is the counterterm
for the corresponding process for a free electron.
The computation of $\Delta E_L = Re(\Delta E_n)$ is quite analogous to the
usual metric case and, although quite laborious, is straightforward.
Working to lowest order in $\varepsilon=1-(\frac{c_*}{c_0})^2$
we find
\begin{equation}\label{el}
\Delta E_L={\cal E}_L\frac{H_0^{1/2}}{c_0^4\epsilon_0^5}
\{p_{n,l}+\varepsilon \tilde p_{n,l}\}
\end{equation}
where
\begin{equation}\label{mp2}
p_{n,l}=
\left\{
\matrix{19/30+
ln(c_0^2\epsilon_0^2/Z\alpha<E_{n,0}>)&\qquad l=0\cr
&\cr
3C_{j,l}/8(2l+1)+
ln(Z^2Ryd/<E_{n,l}>)&\qquad l\not=0\cr}\right.
\end{equation}
\begin{equation}\label{mp3}
\tilde p_{n,l}=-e_{n,l}/8+3p_{n,l}/2 + \frac{183}{80}\delta_{l,0}
\end{equation}
and the reference energy $<E_n>$ is defined as in the usual case \cite{IZ}
by
\begin{equation}
\ln <E_{n,0}>=\frac{\sum_r|<r|{\bf p}|n>|^2(p_0^{(r)}-p_0^{(n)})
\ln(|p_0^{(r)}-p_0^{(n)}|c_0/Ryd^*)}
{\sum_r|<r|{\bf p}|n>|^2(p_0^{(r)}-p_0^{(n)})}
\end{equation}
\begin{equation}
2(\frac{Z\alpha}{c_0\epsilon_0})^4\frac{m^3}{n^3}\ln\frac{Z^2Ryd}{<E_{n,l}>}
= -
\sum_r|<r|{\bf p}|n>|^2(p_0^{(r)}-p_0^{(n)})\ln (c_0|p_0^{(n)}-p_0^{(r)}|)
\end{equation}
with
${\cal E}_L=\frac{4\alpha}{3\pi}\frac{(Z\alpha)^4}{n^3}m_e$,
$e_{n,l}={a_0^*}^3<n|\frac{{\bf x}\cdot{\bf\nabla}}{|{\bf x}|^3}|n>$,
$a_0^*=\frac{\epsilon_0c_0}{H_0^{1/2}}(Z\alpha m_e)^{-1}$ and
$Ryd^*=\frac{H_0^{1/2}}{c_0\epsilon_0^2}Ryd$ being the Rydberg energy.

The physical implications of (\ref{el}) are most easily parametrized in
terms of anomalous mass tensors \cite{Haug}.
A breakdown of
the EEP is signalled by the position and velocity dependance of the binding
energy $E_B$ of a body; this has the general form
\begin{equation}\label{g3}
E_B({\bf X},{\bf V})=E_B^0+\delta m_P^{ij}U^{ij}-\frac{1}{2}\delta
m_I^{ij}V^iV^j + \ldots
\end{equation}
where ${\bf X}$ and ${\bf V}$ are quasi-Newtonian coordinates and velocity
of the
center of mass of the body.  Here $\delta m_P^{ij}$
 and $\delta m_I^{ij}$ are the anomalous
passive gravitational and inertial mass tensors respectively, whose form
depends upon the detailed internal structure of the composite body.
Here
$U^{ij}$ is the external gravitational potential tensor, which satisfies
$U^{ii}=U$.

Violations of LPI occur whenever $\delta m_P^{ij}\neq 0$.
Taking the composite body to be a hydrogen-like atom, from (\ref{el})
we have calculated this to be
\begin{equation}\label{mp}
\delta m_P^{ij(L)}=\delta^{ij}\frac{E_B^L}{c_0^2}(A_{n,l}\Gamma_0+
B_{n,l}\Lambda_0) \equiv \delta^{ij}\alpha_{\mbox{Lamb}}
\ee
and lead to a redshift $\Delta z = (1+\alpha_{\mbox{Lamb}})\Delta U$.
Here
\be
A_{n,l}=5-B_{n,l}-2\delta_{l,0}
(1+3\varepsilon/2p_{n,l})
\qquad B_{n,l}=r_{n,l}(1-\varepsilon(1+r_{n,l}))
\ee
and $r_{n,l}=\frac{\tilde p_{n,l}}{p_{n,l}}$ and
$E_B^L =
-\frac{{\cal E}_L}{(c_0\epsilon_0)^5}(p_{n,l}+\varepsilon \tilde p_{n,l})$
and the quantities $\Gamma_0$ and $\Lambda_0$ are \cite{Haug,WillBanf}
\be
\Gamma_0=\frac{2T_0}{T_0'}(\frac{\epsilon_0'}{\epsilon_0}
+\frac{T_0'}{2T_0}-\frac{H_0'}{H_0})\qquad
\Lambda_0=\frac{2T_0}{T_0'}(\frac{\mu_0'}{\mu_0} +\frac{T_0'}{2T_0}-
\frac{H_0'}{H_0}) \quad .
\ee
It is an experimental challenge to design a
gravitational redshift experiment using a clock sensitive to Lamb shift
transition frequencies. Such an experiment would provide new observational
constraints on the allowed regions of $\Gamma_0$--$\Lambda_0$ parameter
space, enhancing our understanding of the foundations of gravitational
theory.

A computation of $\delta m_I^{ij}$ is in progress and will be
reported upon in the near future \cite{progCat}.

\noindent{\bf Acknowledgements}$\quad$
This work was supported by the Natural Sciences
and Engineering Research Council of Canada. We are grateful to M. Haugan
for discussions. One of us (R.B.M.) would like to thank C.M. Will for a
discussion at  the GR13 meeting in Cordoba where he suggested this problem.

\end{document}